# Measuring Fit of Sequence Data to Phylogenetic Model: Gain of Power using Marginal Tests

Peter J. Waddell[1,2,3], Rissa Ota[2], and David Penny[2]

pwaddell@purdue.edu

[1] SCCC, USC, Columbia, SC 29203, U.S.A.,
[2] Allan Wilson Center for Ecology and Evolution, Massey University, Palmerston North, New Zealand
[3] Currently, Department of Biological Sciences, Purdue University, West Lafayette, IN 47906, U.S.A.

Testing fit of data to model is fundamentally important to any science, but publications in the field of phylogenetics rarely do this. Such analyses discard fundamental aspects of science as prescribed by Karl Popper. Indeed, not without cause, Popper (1978) once argued that evolutionary biology was unscientific as its hypotheses were untestable. Here we trace developments in assessing fit from Penny et al. (1982) to the present. We compare the general log-likelihood ratio (the G or $G^2$ statistic) statistic between the evolutionary tree model and the multinomial model with that of marginalized tests applied to an alignment (using placental mammal coding sequence data). It is seen that the most general test does not reject the fit of data to model (p~0.5), but the marginalized tests do. Tests on pair-wise frequency (**F**) matrices, strongly (p < 0.001) reject the most general phylogenetic (GTR) models commonly in use. It is also clear (p < 0.01) that the sequences are not stationary in their nucleotide composition. Deviations from stationarity and homogeneity seem to be unevenly distributed amongst taxa; not necessarily those expected from examining other regions of the genome. By marginalizing the $4^t$ patterns of the i.i.d. model to observed and expected parsimony counts, that is, from constant sites, to singletons, to parsimony informative characters of a minimum possible length, then the likelihood ratio test regains power, and it too rejects the evolutionary model with p << 0.001. Given such behavior over relatively recent evolutionary time, readers in general should maintain a healthy skepticism of results, as the scale of the systematic errors in published analyses may really be far larger than the analytical methods (e.g., bootstrap) report.

**Keywords**: Fit of sequence data to evolutionary model, base composition stationarity, placental / eutherian mammals.

# 1 Introduction

Measuring the fit of data to model is a fundamental principal in statistics and science (e.g., Popper 1978). It allows, in principal, the data to reject the model being imposed on it, and this in turn warns that any inferences drawn from the model might not only be incorrect, but "positively misleading" (e.g. Felsenstein 1978). This may be interpreted as hyperbole, but it may also be read as capable of yielding strongly supported false positive statistical results. Indeed, when statisticians encounter a clear mismatch of data to model, they are very reluctant to draw firm conclusions about the nature of reality.

This is such a fundamental principal of the statistical sciences, that statisticians are deeply concerned about the power or sensitivity of such tests. Inferring evolutionary trees from real data is best treated as a statistical problem (Felsenstein 1982). While it is known that, asymptotically, tests such as the log likelihood ratio test of a specific model to the general multinomial model are most uniformly powerful, it is important to examine their performance in given instances (e.g. McCullagh and Nelder 1989). Sequence alignments are just such a case since the number of site patterns grows at $4^t$ (where $t$ is the number of taxa), the number of observed site patterns is typically $<10^4$, and the probability of site patterns is highly unequal. This log likelihood-ratio test goes by many names; it is called the G-test, and when multiplied by two so that its asymptotic distribution follows a $\chi^2$ distribution, then it is called a $G^2$ test (i.e., $\ln[G^2] = 2\ln[G]$). It is also equal to twice the entropy difference of two sets of numbers and is the Kullback-Leibler information difference of two distributions (e.g., Sokal and Rolf 1994).

Penny et al. (1982) helped to raise concerns about the issue of whether evolutionary models could reject the data. Indeed, as they mentioned, Popper for a long time regarded historical sciences, such as phylogenetics, as "non-scientific" as the conclusions are untestable and hence unfalsifiable. The analyses in Penny et al. tested a specific prediction of evolutionary theory, namely, that evolutionary trees of different genes should be more similar than expected by chance. Applications of general fit statistics between data and evolutionary model followed. They included Bulmer (1991) using a Mahalanobis or GLS distance between observed and expected (tree) distances. Reeves (1992) and Goldman (1993) used the G-test on sequences and used simulations to account for any bias due to *a posteriori* selection of the tree and due to sparseness of the data. Adachi and Hasegawa (1996) and Waddell (1995) explored use of $X^2$ statistics in place of $G^2$ to measure general fit. These have the advantage of clearly showing which site patterns have the most surprising deviations from expected values. Waddell (1995) compared Mahalanobis distances on phylogenetic distances and spectra, and showed the later is asymptotically equivalent to the $G^2$ test. Also, a resampling approximation of the $G^2$ test was shown to be fast and conservative, yet still capable of rejecting fit of data to model. Spectral analysis and the Hadamard conjugation are also important tools in gauging fit (e.g., Hendy and Penny 1993, Waddell et al. 1997), and these too have lead to surprising findings. One example lead to the identification of strong evidence for extensive ancestral polymorphism within a contiguous stretch of DNA and the estimates of the ancestral population size of humans and chimps (Waddell 1995).

Fit tests based on various marginalization's of the character patterns, have also been proposed. Tavaré (1986) introduced marginal tests of base composition between pairs of taxa, a test of the marginal symmetry of the pair-wise divergence matrix. Rzhetsky and Nei (1995) extended this test to three of more taxa. Waddell and Steel (1996, 1997) showed how the pairwise divergence or **F** matrix is expected to be symmetric under a wide range of time reversible models, and all clock-models, including allowing for unequal rates across sites and allowing invariant sites to have their own frequency distribution. Waddell et al. (2005) tested sequence data at the nucleotide, amino acid, and codon level summing non-intersecting paths **F** tests of symmetry and marginal symmetry. Advantages of this approach over use over the direct extension of tests of symmetry to three or more taxa (Rzhetsky and Nei 1995, Ababneh et al. 2006) include complexity, computability, speed, ready identification of deviant sequences and straightforward grouping to reduce sparseness and better approximate a $\chi^2$ distribution under the null model. Other summaries of such tests are available including Waddell et al. (2005) and Jermiin et al. (2008). It is beneficial to consider both articles, as they tend to focus on different results and/or sets of authors.

Molecular biologists are perhaps the major inferers of evolutionary trees. When faced with difficult intellectual problems, their solutions are often novel. In the field of phylogenetics they often present their work with various ancillary statistics such as bootstrap proportions or posterior probabilities, to argue the importance of their results. Their solution to the problem of fit of data to model has been almost unanimously to ignore it. This approach also 'solves' a second potential issue, that a problem should not be considered resolved when the data do not fit the model. It may argued that much of what passes for clear results in phylogenetics is often neither statistically solid nor science, as Popper (1978) would define it. Indeed, what are often presented as conclusive statistical analyses in phylogenetics is quite possibly indistinguishable from data mining (Waddell and Shelly 2003).

In our own analyses, we have tried to follow good science; for example, extensive analyses on resolving the relationships of the orders of mammals showed the data were seriously compromised by lack of fit of data to model (e.g., Waddell et al. 1999a). Accordingly, while there was repeated congruence of results, and results that only made theoretical sense if certain groups were monophyletic, it was inappropriate to simply present algorithmic outputs. Given a lack of fit of data to model, the appropriate statistical and scientific approach is to conclude that the analyses remain exploratory and using expert opinion, set up the most favored hypotheses for future testing. An example of this is the results of Waddell et al. (1999b). Such *a priori* hypotheses should then be compared with the results obtained from independent data and, if this data fits a model, it may be used to construct an unbiased test the prior hypotheses. An example of such testing, using independent data, is given in Waddell et al. (2001). Here a new statistical test and SINE data were combined to test an *a priori* hypothesis (Zietkiewicz et al. 1999, with much earlier work by Goodman and colleagues, e.g., Goodman et al. 1990). Corroborative approaches, using sequence data (e.g., Lin et al. 2002) have also proven useful. Alternative approaches, including data mining-style analyses, have, in contrast, mislead readers and turned out to be far less accurate. Recent evidence for example, is supporting the clade Atlantogenata (e.g., Waddell et al. 2006, Waters et al. 2007), over alternatives put forward by both morphologists and molecular biologists. A significant contribution to the accuracy of these results was a robust understanding of the importance of fit of data to model, and being open about when results were potentially misleading.

In this work we look at the power of various fits of phylogenetic data to model, including the robust simulations of Reeves (1992) and Goldman (1993), the fast approximation of Waddell (1995), and various intermediaries. These are made upon a subset of the data presented in Waddell and Shelly (2003) to test the tree of Waddell et al. (1999b) against alternatives. We then consider what the tests of Waddell et al. (2005) suggest. Finally, we present a new marginalization of the data that seems to improve the power of the general test, at least on this data.

## 2  Materials and Methods

The RAG1 sequence alignment of Waddell and Shelly (2003) was edited so that there was no missing data or gaps and all positions appeared well aligned (available upon request). This involved the removal of the partial tree shrew sequence, which appeared quite divergent from other species (Waddell et al. 2005). Constraints followed groups for which there is considerable data, in addition to sequences, corroborating the relationships seen in the trees of Waddell et al. (1999b, 2001). These include LINE data (e.g., Kriegs et al. 2005), chromosome rearrangements (e.g., Robinson et al. 2005), and indels (e.g., Waddell et al. 2006). Parts of the tree left unresolved included a three way split at the base of Cetartiodactyla, and the relationships of orders within Scrotifera, Paenungulata and Afroinsectivora.

The constrained ML tree under the GTR invariant sites plus $\Gamma$ model was found using successive iterations of tree search then repotimizing the parameters on the best tree until no further improvement was found using PAUP* (Swofford 2000). Parametric replicate data sets were generated using Seq-gen (Rambaut and Grassly 1997) and analyzed sequentially by PAUP* to give the log likelihood ratio statistic under various scenarios. Results were tabulated using a C-program written by R.O. and plotted using Excel. FreqNuc from Interrogate (Waddell et al. 2005) was used to perform tests on pairwise F matrices. Visualization of results used hierarchical clustering (UPGMA) and Fitch-Margoliash least squares tree fitting excluding negative edges (FM+) as provided by the program PAUP* (Swofford 2000). To calculate the expected frequency over all the $4^t$ possible patterns of sites with a parsimony length of $0 - x$ showing particular nucleotides, we simulated a long sequence ($\sim 10^6$) using ML parameters and Seq-Gen then counted the frequency of sites having observed parsimony properties using PAUP and a C-program written by R.O.

# 3 Results

## 3.1 Fit of RAG1 to a tree using $G^2$

PAUP* was used to find an ML tree using the GTR invariant sites $\Gamma$ model (Waddell and Steel 1997) as shown in appendix 1. This model is an extension of the K3P invariant sites $\Gamma$ model (Waddell and Penny 1996, site pattern probabilities from Steel et al. 1993) used to assess fit of data to model in Waddell (1995). A constraint tree (appendix 1) was used to enforce clades that are supported by both molecular data and rare genomic events such as LINE insertions. Note that a constraint tree should increase the power of the test if it is correct, as it can only increase the log likelihood difference between the ML tree and the general model. If it is wrong, it should bias the test towards rejecting the fit of data to model. Lack of a constraint tree, and propensity to pick an incorrect tree, is more problematic as it may bias the test towards accepting the null hypothesis of model fitting the data when this is not true.

Based on the ML parameters of this evolutionary model, 1000 parametric bootstrap samples were created using the program Seq-Gen. The likelihood of the unconstrained model is calculated using the observed frequency of each site pattern on each replicate. Note, if the data had not been edited to remove sites with indels, then methods like those in Waddell (2005) are required to calculate this unconstrained likelihood.

The likelihood of tree models was calculated using four successively more computationally intensive methods. These were:

> The approximation of Waddell (1995) fixing the tree and ML parameters to those found for the original data.
>
> Reoptimization under the GTR-invariant sites G model on the original ML tree (i.e., no_tree_search).
>
> Reoptimizing the GTR invariant sites parameters after a constrained NNI search (a suggestion by Goldman 1993a).
>
> A constrained TBR search representing a full tree search (Reeves 1992, Goldman 1993a).

We also evaluated the effect of fast approximate tree search using neighbor-joining (Saitou and Nei 1987). For the NJ_JC method, the original $G^2$ was calculated by estimating JC distances and finding the NJ tree consistent with the constraints, then optimizing all tree parameters of the GTR invariant sites $\Gamma$ model on that tree. Each of the simulated datasets was analyzed in the same way. The same was done for the NJ_GTR method, except that the GTR distances used the original ML parameters (including invariant sites and a $\Gamma$ distribution), and after the constrained NJ tree was calculated all parameters were reoptimized on that tree.

The ML tree has some expected and a few unexpected resolutions of the unconstrained parts of the tree, particularly the relationships of the orders within Scrotifera. However, we found a negligible difference in the test results if we fixed the whole tree *a priori* to that in Waddell (2001), which seems reasonable since the internal edges resolving these parts of the tree where near zero length.

The results are surprising. All the test procedures give very similar p values of close to 0.5. In particular, the test of Waddell (1995) gives nearly identical results to that of the far more computationally expensive routines of Reeves and Goldman. The approximate test in was devised on the expectation that it is the extreme sparseness of the data that leads to nearly all of the distortion between the distribution of $G^2$ with typical data and its asymptotic distribution. Similarly, the test fails to reject the null hypothesis that these sequences evolved according to the GTR invariant sites Γ model. Indeed, the p-value is nowhere near leading to rejection. This is somewhat surprising, since subsets of the data have lead to rejection of the model, and it is not clear to what extent sparseness (more taxa for the same number of sequences) is causing such a result.

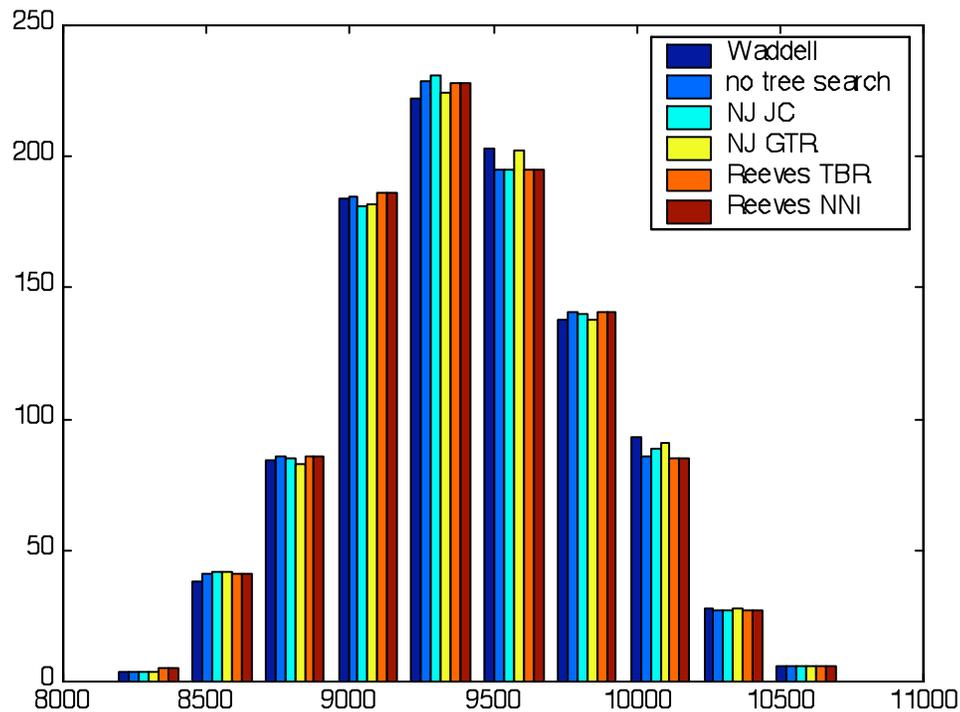

Figure 1. The $G^2$ test between tree model and unconstrained multinomial model for RAG1 data. The original $G^2$ statistic was 9,389 for all but the NJ based tests. For these, the statistic was 9,396 and 9,395 for the JC and GTR based distances, respectively. The bars indicate the $G^2$ statistics from 1000 random samples of data for different models. The boundaries of the categories occur at 8250, 8500, 8750, 9000 and so on. The arrow indicates the approximate position of the observed statistic in relation to the simulated data. The p-values were 0.480, 0.485, 0.489, 0.486, 0.486, and 0.486 for the Waddell, no_tree_search, NJ_JC, NJ_GTR, Reeves_TBR and Reeves_NNI variants of the test, respectively.

All the distributions in figure 1 appear symmetric and close to normal. Under asymptotic conditions, and under the null model and with the tree known *a priori*, a $G^2$ test should converge to a chi-square distribution with degrees of freedom equal to $4^t$-p, where *t* is the number of tips or taxa and *p* the number of parameters estimated in the evolutionary model (McCullagh and Nelder 1989). Here, however, the mean of the actual distribution appears close to 9400. This is clearly much less than $4^{40}$, but it is also considerably larger than the number of sites (730) or site patterns (420) in the observed data. Evaluating the distribution of the test of Waddell (1995) for normality on a larger sample of 3000 draws, using the sensitive test of Anderson and Darling (1952), gives test statistics 0.745 and 0.794 respectively for the W95 and no tree search statistics (p=0.05 critical value = 0.752). However, using *q-q* plots (not shown) there is no clear deviation from normality, but a hint that the upper and lower tails are fatter than expected, suggesting a slightly leptokurtotic distribution. This is encouraging, as it suggests, that for at least some data sets, a reasonable estimate of large sample deviations may be made modeling from a normal distribution.

**3.2 Fit of RAG1 sequence using marginal tests**

Mathematical results in Waddell and Steel (1997) show that under any time reversible model, with any distribution of site rates, and allowing invariant sites to have their own base frequencies, the pairwise divergence matrices are symmetric in expectation (extending earlier results for the equal site rate GTR model of Tavaré 1986). This in turn leads to pairwise likelihood ratio or $X^2$ tests of symmetry (Waddell and Steel 1996, Waddell et al. 1999a, Waddell et al. 2005). With long sequences, this pairwise test statistic for nucleotides is distributed as $\chi^2_6$, that is, chi-square with degrees of freedom (*k*) equals 6. Thus, the mean value is 6, the mode is $k-2 = 4$ (when $k \geq 2$), and the median is $\sim k-2/3 = 5.33$. The critical value at the 5% level is 12.59, and the 1% level is 16.81. Our observations have been that this test typically leads to rejection of the GTR invariant sites model, even though these tests lack full power by summing over non-intersecting paths through the tree. Further, they lack power near the molecular clock, since in this region, homogeneous non-reversible models also show symmetric **F** matrices (Waddell and Steel 1996). Note that reducing the data to a pairwise **F** matrix is a marginalization with respect to the states shown by all other sequences.

Figure 2a graphically shows the results of all the pairwise $X^2$ tests of symmetry of the evolutionary process for this gene region. The strongest outliers are immediately apparent, and these include *Tenrec, Tarsius, Orycterorpus, Sylvilagus* and *Ochotona*. Given the short edges in the tree, it is possible that such changes in the evolutionary model will result in current methods of tree estimation inferring an incorrect tree with considerable statistical support. A probable example, involving the divergent processes of *Tarsius* and *Tupaia*, relative to *Homo* and *Cynocephalus* is given in Waddell, et al. (2005). Another set of closely related taxa that are divergent are the members of Afroinsectivora, that is, *Tenrec, Amblysomus, Orycterorpus*, and *Macroscelides*, and there should be doubts about how reliable the sequence trees within this group are (e.g, Murphy et al. 2001, Waddell et al. 2001, Waddell and Shelly 2003), not least because they conflict with at least one chromosomal character (Robinson et al. 2005). While the bats are not as extreme, they too show divergences from one another, raising the question of whether sequence-based phylogenies of bats (e.g., Teeling et al., 2000) are fully reliable.

Surprisingly, neither *Mus* nor *Erinaceus* appear divergent, given that on other data sets (Waddell et al. 1999a, Waddell et al. 2001) they are strongly divergent from most other species, and that their inclusion or exclusion can result in very different trees, especially with lesser taxon sampling (Waddell et al. 2001).

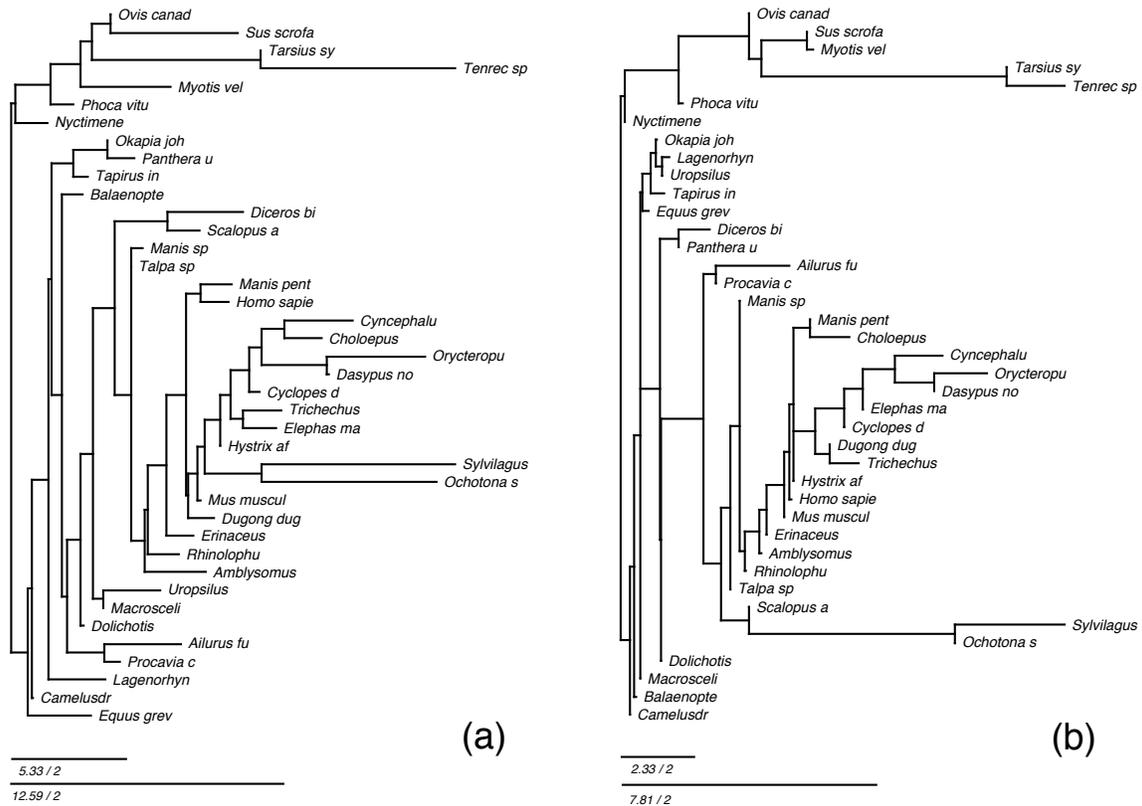

Figure 2. (a) Visualization of all the pairwise $X^2$ tests of symmetry of the divergence (**F**) matrices for the rag1 gene region. The $X^2$ test result was used as a distance, and these were fitted to a tree using the FM+ criterion (% s.d. = 32.65). Half the median and 5% critical values of the test statistic are shown as the scale. (b) As for (a) except the test statistic is the sum of squares from the pairwise GLS test of stationarity statistic.

The evolutionary properties of a set of homologous sequences may be further explored using a test of the stationarity of the evolutionary process. Stationarity may be assessed by looking for significant changes in the marginal frequencies of the nucleotide bases (that is testing for base composition stationarity). Comparing directly the row and the column sums with an $X^2$ or $G^2$ test statistic can do this, but this does not take into account correlations due to some cells of the **F** matrix appearing in different rows and columns. While such correlations may have little effect if applied to amino acid or codon frequencies (Waddell, Hasegawa and Mine 2005), it is prudent to use a GLS type statistic (Tavaré 1986) with just four states. Under the null model, that there is no difference in the expected base frequencies of the evolutionary process along the two lineages linking these species, and with long sequences, this statistic is distributed as $\chi^2_3$, that is, chi-square with degrees of freedom (*k*) equals 3. Thus, the mean value is 3, the mode is *k*–2 = 1 (when $k \geq 2$), and the median is $\sim k$-2/3 = 2.33. The critical value at the 5% level is 7.81, and the 1% level is 11.34.

Figure 2b graphically shows the results of tests of stationarity of the evolutionary process for this gene region. The results are in very good agreement with those of figure 2a based on the reversibility of the model linking pairs of taxa, with relatively minor fine tuning as to exactly how the taxa relate to one another in terms of the directions of their evolution.

A formal test of the hypothesis that sites evolved according to a general time reversible model may be made by summing up the pairwise results of non-intersecting paths through the tree. The results in table 1 are reported by the program FreqNuc (Waddell et al. 2004, 2005). Clearly, the null hypothesis is rejected and there is also clear evidence of changing base composition as well as changes in the patterns of nucleotide changes.

Table 1. Non-intersecting pairs test of the time reversibility and stationarity of the evolution of Rag1 sequences. The pairs of tests are made on non-overlapping paths in the ML tree, and include grouping of cells to improve convergence to asymptotic distributions (Waddell, Hasegawa, and Mine 2005). The p-values for the totals of the $G^2$ tests are ungrouped symmetry $\chi^2_{102}(165.1)$ ~0.00003, grouped symmetry $\chi^2_{62}(101.1)$ ~0.0008, and for the GLS test of stationarity $\chi^2_{51}(78.4)$ ~0.008, respectively. Thus the null hypothesis of time reversible model with unequal site rates is clearly rejected. Also included is an invalid test of stationarity over all sites including constant sites (St.All, Waddell et al. 1999), similar to that still being made in some phylogenetics programs. Its nominal test statistic is $\chi^2_{102}(14.4) \sim 0.99999$, which misleads researchers into reporting their data as consistent with a reversible model. Underlined are the larger pairwise deviations.

| | | Ungrouped | | | Grouped | | | |
| --- | --- | --- | --- | --- | --- | --- | --- | --- |
| Species | | Symmetry | St.All | SS[a] | Symmetry | df[b] | SS5[c] | df[b] |
| *Ovis_canadenensis* | $X^2$ | 4.0 | 0.5 | 2.5 | 3.5 | 4 | 2.5 | 3 |
| *Lagenorhynchus_obsc* | $G^2$ | 4.1 | 0.5 | | 3.5 | 4 | | 3 |
| *Sus_scrofa* | $X^2$ | 8.8 | 0.7 | 3.7 | 8.2 | 4 | 3.7 | 3 |
| *Camelus_dromedarius* | $G^2$ | <u>11.2</u> | 0.7 | | 9.2 | 4 | | 3 |
| *Diceros_bicornis* | $X^2$ | 6.0 | 0.1 | 3.1 | 0 | 2 | 3.1 | 3 |
| *Tapirus_indicus* | $G^2$ | 8.0 | 0.1 | | 0 | 2 | | 3 |
| *Equus_grevyi* | $X^2$ | <u>15.1</u> | 0.4 | <u>3.7</u> | <u>10.3</u> | 3 | <u>3.7</u> | 3 |
| *Rhinolophus_creaghi* | $G^2$ | <u>16.9</u> | 0.4 | | <u>11.1</u> | 3 | | 3 |
| *Phoca_vitulina* | $X^2$ | 3.4 | 0.1 | 2.4 | 0.6 | 2 | 2.4 | 3 |
| *Ailurus_fulgens* | $G^2$ | 4.2 | 0.1 | | 0.6 | 2 | | 3 |
| *Panthera_uncia* | $X^2$ | 4.7 | 0.8 | 4.6 | 3.8 | 3 | 4.6 | 3 |
| *Manis_pentadactyla* | $G^2$ | 4.8 | 0.8 | | 3.8 | 3 | | 3 |
| *Myotis_vellifer* | $X^2$ | 7.0 | 0.5 | 3.6 | 3.4 | 3 | 3.6 | 3 |
| *Nyctimene_albivente* | $G^2$ | 8.0 | 0.5 | | 3.4 | 3 | | 3 |
| *Erinaceus_europaeus* | $X^2$ | 6.1 | 0.7 | 3.6 | 2.5 | 5 | 3.6 | 3 |
| *Cyncephalus_volans* | $G^2$ | 7.3 | 0.7 | | 2.5 | 5 | | 3 |
| *Scalopus_aquaticus* | $X^2$ | 5.1 | 0.5 | 4.0 | 3.7 | 4 | 4.0 | 3 |
| *Uropsilus_sp* | $G^2$ | 5.7 | 0.5 | | 3.7 | 4 | | 3 |
| *Homo_sapiens* | $X^2$ | <u>21.6</u> | 3.3 | <u>15.4</u> | <u>13.2</u> | 4 | <u>15.4</u> | 3 |

| | | | | | | | |
|---|---|---|---|---|---|---|---|
| *Tarsius_syrichta* | $G^2$ | <u>25.9</u> | 3.3 | | <u>13.7</u> | 4 | | 3 |
| *Mus_musculus* | $X^2$ | 3.7 | 0.3 | 1.0 | 2.9 | 5 | 1.0 | 3 |
| *Dolichotis_patagonu* | $G^2$ | 3.9 | 0.3 | | 2.9 | 5 | | 3 |
| *Sylvilagus_sp* | $X^2$ | 9.5 | 0.1 | 1.8 | 1.0 | 3 | 1.8 | 3 |
| *Ochotona_sp* | $G^2$ | 12.3 | 0.1 | | 1.0 | 3 | | 3 |
| *Amblysomus_hottento* | $X^2$ | <u>26.8</u> | 3.8 | <u>14.5</u> | <u>26.8</u> | 6 | <u>14.5</u> | 3 |
| *Tenrec_sp* | $G^2$ | <u>28.5</u> | 3.8 | | <u>28.5</u> | 6 | | 3 |
| *Macroscelides_sp* | $X^2$ | <u>9.3</u> | 1.9 | <u>8.4</u> | <u>9.3</u> | 5 | <u>8.4</u> | 3 |
| *Orycteropus_afer* | $G^2$ | <u>9.8</u> | 1.9 | | <u>9.8</u> | 5 | | 3 |
| *Dugong_dugon* | $X^2$ | 3.7 | 0.3 | 2.6 | 3.4 | 3 | 2.6 | 3 |
| *Procavia_capensis* | $G^2$ | 3.7 | 0.3 | | 3.5 | 3 | | 3 |
| *Cyclopes_didactylus* | $X^2$ | 4.5 | 0.3 | 2.6 | 1.5 | 3 | 2.6 | 3 |
| *Choloepus_hoffmani* | $G^2$ | 4.8 | 0.3 | | 1.6 | 3 | | 3 |
| *Dasypus_novemcinctu* | $X^2$ | 5.1 | 0.2 | 0.8 | 2.1 | 3 | 0.8 | 3 |
| *Elephas_maximus* | $G^2$ | 6.3 | 0.2 | | 2.1 | 3 | | 3 |
| Overall | $X^2$ | 144.4 | 14.4 | 78.4 | 96.1 | 62 | 78.4 | 51 |
| | $G^2$ | 165.1 | 14.4 | | 101.1 | 62 | | 51 |

[a] GLS test of stationarity comparing row versus column sums sum of squares

[b] degrees of freedom of the preceding test which groups so that no expected value is less than 1 and no more than 20% of expected values are less than 5

[c] GLS test of stationarity comparing row versus column sums after grouping so that all expected values are five or greater sum of squares

[d] degrees of freedom of the preceding test

It is interesting to note that much of the violation of the model detected in these sequences is coming from a few pairs of taxa, which are underlined in the table. The strongest effected in table 1 are the pair *Tenrec* and *Amblysomus* (golden mole). Examining figure 2 suggests that *Tenrec* is the main offender here, while *Amblysomus* is moving away, from the average in another direction, but not nearly as strongly. Such affects may partly explain why the group Tenrecomorpha (also called Afrosoricida) has shown weak support in some studies (Waddell and Shelly 2003). Another strongly divergent pair of sequences is *Homo* and *Tarsier*. Looking at figure 2, the main offender here appears to be *Tarsier*, although *Homo* and other relatives such as *Cynocephalus* (flying lemur) are moving towards another extreme. Such problems may underlie the difficulties of locating *Tarsier* sequences relative to other euarchontan species (Waddell et al. 2001). Next, consider the pair *Equus* (horse) and *Rhinolophus* (a bat). Here there is evidence of a different pattern of substitution but not of changing base composition. Figure 2 shows both these taxa to be fairly average, raising the issue as to whether either is a strong offender of overall assumptions. If they are, it is not enough to stop bats and perissodactyls grouping in this gene tree analysis, a clade consistent with suggestions by Okada et al. (2006). A similar result holds for the pair *Sus* (pig) and *Camelus* (camel), and here too there should be suspicion of the relationship of these two to Cetruminatia, based only on sequence data.

A surprising aspect of these tests, and one touched upon earlier in Waddell et al. (2005), are specific taxa that do not show strong deviations from the average. In particular, murid rodents such as *Mus* (mouse) and the eulipotyplan hedgehog (*Erinaceus*), which depart strongly from average in other analyses (e.g. Waddell et al. 1999b, 2001) seem quite average here. This suggests there may be strong temporal and spatial deviations from average in the evolution of mammalian genomes.

**3.2 Recovering power in the likelihood ratio test via marginalization**

We now consider a second type of marginalization. Here, cells (site patterns) will be grouped *a priori* according to their nucleotides (e.g., A and G only) and according to how many parsimony changes they require to fit the tree. The table starts out by comparing observed and expected frequencies of the constant sites. It then progresses on to all possible types of singletons. A singleton is defined as an aligned site at which all but the most common state appears at most once, that is, only one taxon shows the unique state. For example, a singleton may be composed of AG, where A is the more frequent and G the unique (singly occurring) state. There are 12 such unlabeled patterns possible. By unlabeled we mean that if the taxa are unlabeled or undistinguished, there remain 12 patterns that may still be differentiated. A singleton may also be composed of ACG, as long as both the G and the C occur in a sequence once each. There are six such possible patterns, and there are also four unlabeled singleton patterns of the form ACGT. Generally, such states are fairly common in the data, but even then, the less frequent may need to be grouped together in order to come close enough to asymptotic conditions not to be too concerned about the need for another level of simulation (as explained below). Finally, it is possible to group all the parsimony informative sites using the properties of minimum length and their actual length on the tree. Given the necessity of grouping states and also for simplicity, for these we do not discriminate if a site shows any particular combination of A, C, G and T.

After grouping the tails of the distribution to meet standard guidelines for achieving a good approximation to asymptotic conditions, the results of this marginalization are shown in table 2. The overall $X^2$ statistic is over three times as large as expected and highly significant. An examination of cells highlights some of the reasons for this. First, the frequency of C and T at the constant sites, does not agree with the ML estimates. A similar problem seems to occur where there are fewer TC and more AC singletons than expected. However, the most substantial deviations from expected occur amongst the parsimony informative sites. In particular, states showing only two states (two different nucleotides) are showing a lot more multiple changes (homoplasy) than expected. In contrast those showing three states tend to show a deficit of multiple changes, while those with four states also show a less pronounced deficit of changes. Clearly ML is attempting to balance everything out, but the data is not fitting the model and it has left this telling pattern. The exact cause could be a number of things, and this is an issue best addressed by comparing the fit in and with biologically more realistic models, including codon models.

In order to calculate the expected frequencies of sites with higher parsimony counts, while avoiding summing up the $4^t$ possible site patterns, we simulated 100,000 sites under the model. We can calculate approximately how much affect this sampling will have on the overall $X^2$ statistic. Let Y be a sample of size 730 from Z, the true values, while X is a sample of 100000 from Z. If the deviation of Y from Z has expected value of 38, for example, then the expected deviation of X from Z has an expected value of $38/(100000/730)^{0.5}$ or just 3.2 or about 3.2/(38+3.2) or ~8% of the total. This was corroborated by another simulation of 100,000 sites giving a nearly identical result in terms of the overall fit.

Table 2. Marginalization by mapping down to the most frequent patterns based on nucleotide state and number of changes as estimated on the constrained GTR $p_{inv}$ + $\Gamma$ ML model. In this case we are using the $X^2$ and not the $G^2$ statistic since it is useful to examine the contributions of individual cells. There are 39 cells, and since they sum to the sequence length, there are nominally 38 degrees of freedom. A $\chi^2_{38}(122.91)$ ~ 7 × 10$^{-11}$, so the data clearly do not fit the model. This fit is ignoring the $2^{t-3}$ + 8 + 2 parameters fitted to the data in the original fitting, making this a conservative test. Underlined are the larger individual cell deviations.

| Number of states | Type | States | Observed | Expected[a] | $X^2$ |
|---|---|---|---|---|---|
| Parsimony uninformative | | | | | |
| 1 state | | A | 79 | 71.7 | 0.74 |
| | | G | 70 | 76.4 | 0.53 |
| | | C | 55 | 78.7 | 7.15 |
| | | T | 78 | 55.2 | 9.38 |
| 2 states | Purines | AG | 18 | 14.7 | 0.74 |
| | | GA | 11 | 17.7 | 2.54 |
| | Pyrimidines | CT | 18 | 16.1 | 0.22 |
| | | TC | 6 | 13.1 | 3.85 |
| | Mixed | AC | 11 | 5.8 | 4.66 |
| | | GT | 2 | 2.8 | 0.23 |
| | | CA | 7 | 5.3 | 0.55 |
| | | remainder | 16 | 11.8 | 1.49 |
| 3 states | | CGT | 5 | 2.7 | 1.96 |
| | | remainder | 15 | 22.2 | 2.34 |
| 4 states | All | | 0 | 3.4 | 3.40 |
| Parsimony informative | | Length[b] | | | |
| Minimum length = 1 | | 1 | 21 | 22.8 | 0.14 |
| | | 2 | 65 | 48.0 | 6.02 |
| | | 3 | 34 | 25.5 | 2.83 |
| | | 4 | 25 | 12.9 | 11.35 |
| | | 5 | 15 | 6.6 | 10.69 |
| | | 6 | 7 | 3.1 | 4.91 |
| | | 7+ | 11 | 2.5 | 28.90 |
| Minimum length = 2 | | 2 | 12 | 15.4 | 0.75 |
| | | 3 | 31 | 38.2 | 1.36 |
| | | 4 | 16 | 29.7 | 6.32 |
| | | 5 | 19 | 21.8 | 0.36 |
| | | 6 | 8 | 14.4 | 2.84 |
| | | 7 | 11 | 9.7 | 0.17 |
| | | 8 | 4 | 5.7 | 0.51 |
| | | 9 | 3 | 3.2 | 0.01 |
| | | 10+ | 4 | 2.8 | 0.51 |

| | | | | |
|---|---|---|---|---|
| Minimum Length = 3 | 3-4 | 10 | 13.4 | 0.86 |
| | 5 | 9 | 11.4 | 0.51 |
| | 6 | 6 | 11 | 2.27 |
| | 7 | 7 | 10.3 | 1.06 |
| | 8 | 6 | 8.2 | 0.59 |
| | 9 | 6 | 6.5 | 0.04 |
| | 10 | 4 | 4.3 | 0.02 |
| | 11+ | 4 | 4.7 | 0.10 |
| | | | Sum | 122.91 |

[a] These expected values are estimated from 100000 sites randomly generated by SeqGen

[b] Unweighted parsimony length on the constrained ML tree

## 4 Discussion

It is interesting that while the general $G^2$ likelihood ratio test lacks power, the marginal tests do detect clear deviations from the model. Further, the different marginalizations seem to be detecting different deviations. The pairwise **F** matrices are showing up taxa that are changing in their properties, while grouping parsimony patterns suggests excesses and deficiencies in the observed and expected number of changes. It is however possible that these two are interrelated, due to the ML model struggling to find a happy medium between a majority of taxa that are fairly close to the model and a few that are further away.

While it is seen here that the $G^2$ tests do not reject the model, the reader should not take away the impression that pairwise **F** tests are an alternative. It is a corollary of results in Waddell (1995), Waddell and Steel (1996), and Waddell (1997) that the **F** statistics may not detect some types of compositional drift. This is because under a clock sequences evolving by a non-GTR model will show symmetric **F** matrices. In contrast, GTR models with a specified distribution of site rates are always identifiable; therefore, any deviation from these expectations is detectable given long enough sequences, clock or not. Further, while identifiability may be compromised (Steel et al. 1994), so far the only know cases involve few taxa, special arrangements of edge length, and very short internal edge lengths if path-lengths through the tree are bounded at sensible values (Waddell 1995).

One of the issues with the $G^2$ test is needing it, but not relying upon it too much. On sequences Foster (2004) found it did not identify some data sets where a naïve test of base composition requiring simulations did. Another example where this test was found wanting was in graphical modeling of gene regulatory networks (Waddell and Kishino 2000). However, we need to be cautious not to ignore the $G^2$ test as it is uniformly most powerful test. Base composition is one thing we may be looking for, but a multinomial test is in theory capable of detecting any deviation, for example, signals from multiple gene trees (Waddell 1995). Since the multinomial test is based on a sufficient statistic, then, in theory, all other tests can be constructed from it. It cannot be replaced, but is needs to be supplemented with specific tests of which a range has been developed (e.g., Goldman 1993b, Waddell 1995, Ota et al. 2000) and more are needed.

This research has suggested future directions currently being explored. One is that the calculation of the grouped patterns in table 2 is not efficient and awaits a computationally efficient algorithm, if one exists. It is also desirable to have an efficient algorithm to calculate the maximum weight set of non-intersecting paths using in table 1. Another is evaluating the power and utility of the following test that has the potential to both reject the model and identify parts of the tree that are problematic. Firstly, construct the observed **F** matrix on each edge of an ML tree by summing over all the conditional site reconstructions. The expected edge matrix may be written, $\mathbf{F} = \Pi \phi[\mathbf{R}t]$, where $\Pi$ is a diagonal matrix of the equilibrium frequencies of the ML rate matrix $\mathbf{R}$, $t$ is an edge scalar, and $\phi$ is the inverse of the moment generating function of the site rate variability (Waddell and Steel 1997). Following that each pair of observed and expected **F** matrices may be tested as in Waddell et al. (2004), and the results summed. Finally, it is straightforward to extend the testing procedures of Waddell et al. (2004) to include the relationship noted by Ababneh et al. (2006). That is, the internal symmetry statistic of **F**, $X^{2*}[\mathbf{F}]$, equals $X^2[\mathbf{F}]$ - GLS[**F**].

Tests like these were used extensively to assess the data used in papers such as Waddell et al. (1999b, 2001) and Waddell and Shelly (2003). They clearly show that even nuclear sequences of placental mammals are prone to significant base composition shifts. We have long suspected that the non-stationarity/non-homogeneity of the evolutionary process is largely responsible for groupings that contradict the better morphological data near tips of the tree. Accordingly, we have consistently stuck with morphological hypotheses such as Primatomorpha (Primates and flying lemurs) and Tethytheria (elephants and sea cows) in our favored trees, while other authors have been inclined to present the alternatives as probable. Running these test on other genes have shown similar results, but it is interesting that different genes show different degrees of deviation, and taxa that are problematic in some genes (e.g., *Mus, Erinaceus*), look quite normal in others

Finally, in view of the rarity with which diagnosis of fit of data to model is pursued in phylogenetics, and the continued development of fast and powerful tests to identify inadequacies in evolutionary models, phylogenetics as a field needs to be humble in how strongly they proclaim particular parts of the tree of life resolved. Failure to do this invites ridicule by more mature statistical sciences and ignores the seriousness of Popper's concerns regarding historical subjects being true sciences.

**Acknowledgements**

This work was supported by NIH grant 5R01LM008626 to PJW. Thanks to Mike Steel for helpful discussions.

**Author contributions**

PJW originated the research, developed methods, gathered data, ran analyses, prepared figures and wrote the manuscript. RO programmed summation routines and ran analyses. David Penny provided comments and support via Allan Wilson Center funding.

## Appendix 1

The constraint tree used:

(((((Sus_scrofa,Camelus_dromedarius,((Ovis_canadenensis,Okapia_johnstoni),(Lagenorhynchus_obscurus,Balaenoptera_physalus))),(Equus_grevyi,(Diceros_bicornis,Tapirus_indicus)),(Panthera_uncia,(Phoca_vitulina,Ailurus_fulgens)),(Manis_sp,Manis_pentadactyla),(Rhinolophus_creaghi,(Myotis_vellifer,Nyctimene_albiventer))),(Erinaceus_europaeus,(Uropsilus_sp,(Scalopus_aquaticus,Talpa_sp)))),((Cyncephalus_volans,(Homo_sapiens,Tarsius_syrichta)),((Mus_musculus,(Dolichotis_patagonum,Hystrix_africaeaustralis)),(Sylvilagus_sp,Ochotona_sp)))),(((Macroscelides_sp,Orycteropus_afer,(Amblysomus_hottentotus,Tenrec_sp)),(Procavia_capensis,Elephas_maximus,(Dugong_dugon,Trichechus_manatus))),(Dasypus_novemcinctus,(Cyclopes_didactylus,Choloepus_hoffmani))));

The ML tree used:

((((((Camelus_dromedarius,(Sus_scrofa,((Ovis_canadenensis,Okapia_johnstoni),(Lagenorhynchus_obscurus,Balaenoptera_physalus)))),(Panthera_uncia,(Phoca_vitulina,Ailurus_fulgens))),(((Equus_grevyi,(Diceros_bicornis,Tapirus_indicus)),(Rhinolophus_creaghi,(Myotis_vellifer,Nyctimene_albiventer))),(Manis_sp,Manis_pentadactyla))),(Erinaceus_europaeus,(Uropsilus_sp,(Scalopus_aquaticus,Talpa_sp)))),((Cyncephalus_volans,(Homo_sapiens,Tarsius_syrichta)),((Mus_musculus,(Dolichotis_patagonum,Hystrix_africaeaustralis)),(Sylvilagus_sp,Ochotona_sp)))),(((Orycteropus_afer,(Macroscelides_sp,(Amblysomus_hottentotus,Tenrec_sp))),(Elephas_maximus,(Procavia_capensis,(Dugong_dugon,Trichechus_manatus)))),(Dasypus_novemcinctus,(Cyclopes_didactylus,Choloepus_hoffmani))));